\documentclass[journal]{IEEEtran}
\usepackage{makecell}
\usepackage{multirow}

\usepackage{cite}

\ifCLASSINFOpdf
  \usepackage[pdftex]{graphicx}
  \graphicspath{{../pdf/}{../jpeg/}}
  \DeclareGraphicsExtensions{.pdf,.jpeg,.png}
\else
  \usepackage[dvips]{graphicx}
  \graphicspath{{../eps/}}
  \DeclareGraphicsExtensions{.eps}
\fi
\usepackage{amsmath}
\usepackage{array}
\usepackage{url}

\begin{document}
%
\title{Cross-Referencing Self-Training Network for \\ Sound Event Detection in Audio Mixtures}
%
%
%

\author{Sangwook Park,
        David K. Han,~\IEEEmembership{Senior Member,~IEEE}
        Mounya Elhilali,~\IEEEmembership{Senior Member,~IEEE}%
\thanks{S. Park is with the Department
of Electrical and Computer Engineering, Johns Hopkins University, Baltimore,
MD, 21218 USA.}
\thanks{D. K. Han is with the Department of Electrical and Computer Engineering, Drexel University, Philadelphia, PA, 19104 USA.}
\thanks{M. Elhilali is with the Department
of Electrical and Computer Engineering and jointly with the Department of Psychology and Brain Sciences, Johns Hopkins University, Baltimore,
MD, 21218 USA e-mail: mounya@jhu.edu.}
}

\maketitle

\begin{abstract}
Sound event detection is an important facet of audio tagging that aims to identify sounds of interest and define both the sound category and time boundaries for each sound event in a continuous recording. With advances in deep neural networks, there has been tremendous improvement in the performance of sound event detection systems, although at the expense of costly data collection and labeling efforts. In fact, current state-of-the-art methods employ supervised training methods that leverage large amounts of data samples and corresponding labels in order to facilitate identification of sound category and time stamps of events. As an alternative, the current study proposes a semi-supervised method for generating pseudo-labels from unsupervised data using a student-teacher scheme that balances self-training and cross-training. Additionally, this paper explores post-processing which extracts sound intervals from network prediction, for further improvement in sound event detection performance. The proposed approach is evaluated on sound event detection task for the DCASE2020 challenge. The results of these methods on both "validation" and "public evaluation" sets of DESED database show significant improvement compared to the state-of-the art systems in semi-supervised learning.
\end{abstract}

\begin{IEEEkeywords}
Sound event detection, semi-supervised learning, self-training, pseudo label.
\end{IEEEkeywords}

%
\IEEEpeerreviewmaketitle

\section{Introduction}
\IEEEPARstart{A}{UDIO} tagging summarizes an audio stream with descriptive information pertaining to the sound in terms of the location where the sound may be coming from, emotional content present, causal relationships among the sound sources, or other descriptive information. Aside from the informative nature, they can also be quite helpful in expediently retrieving or categorizing audio as the size of a typical audio database is quite large. Sound Event Detection (SED), which aims to identify sounds of interest in terms of sound category and its temporal boundaries, has been a critical technique for audio tagging. Since sounds are quite informative to understand the auditory scene such as a presence of human, animal, or any entities, and its behavior, the technique has been adopted for many applications including video analytics, baby or pets monitoring, and other surveillance systems~\cite{Ahmad2019, Lavner2017, Park2015}. For performing audio tagging in an environment, an SED method should be able to identify multiple sounds even when these sounds overlap with each other temporally.

With recent advances in deep learning, deep neural networks have shown outstanding improvements in SED~\cite{Cakir2017,Vesperini2019,Kothinti2019,Kong2020}. To train a deep network for a SED task, each training sample needs to be annotated with the sound class and time boundaries of every target sound interval therein. The annotation allows the network to learn spectro-temporal characteristics of the target sound in a supervised fashion. Accurate labels and the markings of temporal boundaries are critical to train the model; however, generating them is often quite expensive and time consuming. Semi-supervised learning, which leverages extensive unlabeled data in combination with small amounts of labeled data, has been explored to resolve the issue in data collection~\cite{Shi2019, Kothinti2019, Lin2020}. In the recent challenge of the Detection and Classification of Acoustic Scenes and Events (DCASE) 2020, task 4 involves building an SED model in a semi-supervised fashion. The task provides an extensive set of unlabeled data as well as a smaller set of weakly labeled data with labels describing the sound class only~\cite{Turpault2020}.

Among semi-supervised techniques, self-training is an intuitive approach that is easy to understand and whose theoretical feasibility has been studied in several works~\cite{Lee2013a,Wei2021}. Self-training effectively uses the prediction of a network for a given input as a pseudo-label to further train the network. As such, the accuracy of this pseudo label has important implications on the network's performance. In an earlier work, a method of pseudo label estimation for unlabeled data and a reliability of the pseudo label were proposed~\cite{Park2021a}. The pseudo label was estimated by a probabilistic expectation of all potential labels as these probabilities were calculated based on the Bernoulli process with posterior probabilities of each class. With labeled data, reliability of the pseudo label at each training step was measured based on a binary cross entropy between true label and the estimate for the labeled data. The objective function was composed of a supervised loss for the labeled data computed from a cross-entropy between the true label and the model prediction, and an expectation loss for the unlabeled data, which was defined as a mean-squared error between the pseudo label and the prediction. The expectation loss was weighted by the reliability. Then, the model is self-trained by performing the estimation and optimization in every training step. 

As an extension to the previous work, the current paper proposes a Cross-Referencing Self-Training (CRST) model. A critical issue with self-training is the self-referencing framework that has a risk of self-biasing due to the pseudo label estimated by itself. To resolve this issue, dual models composed of \textit{Model I} trained with original data and \textit{Model II} trained with perturbed data, are incorporated in the self-training. Each of these models is trained separately using the pseudo-label estimate of the other network. Additionally, this paper explores a post-processing step to extract target sound intervals from the network prediction with a classwise thresholding and smoothing. For thresholding, the Extreme Value Theory (EVT) based threshold estimation is performed for each target class. A smoothing filter length is determined based on statistics of each target sound duration. To demonstrate effectiveness of the proposed method, experiments are performed following the protocol for the multi-target SED task in the recent DCASE challenge (DCASE2020). The result shows further improvement in the performance compared to the previous model as well as other state of the art in semi-supervised learning. The main contribution of this paper can be summarized as: 1) a novel self-training method which avoids the self-biasing issue; 2) an effective approach of selectively combining synthetic data with unlabeled or weakly labeled real world data to enable a semi-supervised training; 3) introducing a classwise post-processing involving effective estimates of thresholds and durations for further improvement in SED performance.

The rest of this paper is organized as follows. Related works for semi-supervised learning are explored in the following section. The motivation of the self-training model is described in Section III, and Section IV describes the proposed methods for the cross referencing self-training model and the classwise post-processing. In Section V, experiment results performed on the DCASE challenge framework are summarized. Comparison to the challenge submission is discussed in Section VI and the conclusion is followed.

\section{Related works}
\subsection{Semi-supervised learning}
Semi-supervised learning aims to leverage unlabeled data to improve the performance of supervised learning with a small labeled data set. There has been a growing body of work exploring use of unlabeled data in supervised learning~\cite{VanEngelen2020}. Usually, unlabeled data is used to learn a preliminary model about the input distribution, and the model is used for either feature extraction~\cite{Mun2017, Mikolov2013, Sindhwani2006}, clustering of data to assign label~\cite{Dara2002}, or initialization of parameters~\cite{Erhan2010}. These methods are then leveraged in a main model for classification which is trained with labeled data in supervised passion.

Particularly, the clustering based approach assumes that two samples belonging to same cluster in observation space are likely to belong to the same class. This is known as the cluster assumption (or low-density separation assumption) and has inspired a consistency-regularization approach like PI-model~\cite{Sajjadi2016, Laine2017}, temporal ensemble model~\cite{Laine2017}, Mean Teacher (MT) model~\cite{Tarvainen2017}, and Interpolation Consistency Training (ICT) model~\cite{Verma2019}. Among these approaches, the MT model has been instrumental in pushing forth the state of the art. The MT model consists of two networks, \textit{student} and \textit{teacher}, and its objective function is denoted as
\begin{equation}
\label{eq:MTloss}
    L_{MT} = BCE(y, f_{\theta}(x)) + \delta MSE(f_{\theta}(x;\eta),f_{\theta'}(x;\eta')),
\end{equation}
where $BCE(y, f_{\theta}(x))$ is a classification loss implemented by a Binary Cross Entropy (BCE) between true label $y$ and \textit{student} prediction $f_{\theta}(x)$. $MSE(f_{\theta}(x;\eta), f_{\theta'}(x;\eta'))$ is a consistency loss implemented by a Mean Squared Error (MSE) between two predictions $f_{\theta}(x;\eta)$ by \textit{student} and $f_{\theta'}(x;\eta')$ by \textit{teacher} under a random perturbation $\eta$ on each network such a rotating, shifting, or adding noise. Network parameters are represented as $\theta$ and $\theta'$ for \textit{student} and \textit{teacher}, respectively. This consistency loss is controlled by the $\delta$ which is usually designed with ramp-up value during the training. Both networks are constructed with the same architecture. The \textit{student} parameters are updated by using a gradient descent method. On the other hand, parameters in the \textit{teacher} model are updated by exponential moving average of \textit{student} parameters over the training step. Since the averaging network tends to produce more accurate prediction than a network obtained by the gradient descent method in each training step~\cite{Tarvainen2017}, the \textit{student} network is guided by the \textit{teacher} network. Additionally, the consistency loss enables that the \textit{student} network produces the same predictions even in presence of various perturbations. Once the MT model converges in training, the \textit{student} network projects any samples belonging to the manifold constructed by the perturbations onto the same prediction.

In contrast, the ICT model differs from the MT model in that an interpolation between two inputs is considered instead of random perturbations. The objective function is denoted as
\begin{equation}
\label{eq:ICTloss}
\begin{aligned}
    L_{ICT} &= BCE(y, f_{\theta}(x)) + \\
    & \delta MSE(f_{\theta}(Mix_{\lambda}(x_1, x_2)), Mix_{\lambda}(f_{\theta'}(x_1),f_{\theta'}(x_2))), \\
    &~where~~Mix_{\lambda}(a,b) = \lambda a + (1-\lambda)b,
\end{aligned}
\end{equation}
where $\lambda$ is a random value as $0 \leq \lambda \leq 1$. The second term enables that the \textit{student} network projects a convex set of the inputs onto another convex set of predictions for the inputs by itself. Once an ICT model converges during training, the \textit{student} network produces similar predictions for any samples in the convex set of the inputs. This is more efficient compared to random perturbations since interpolation always generates a sample belonging to the convex set of the inputs while not all samples generated by random perturbations belong to the manifold for the actual inputs.

As an alternative approach, the present study explores a self-training method which estimates a pseudo label for unlabeled data, then updates itself for both labeled data and pseudo labeled data (unlabeled data). As a simple way to estimate pseudo label, Lee assigned a pseudo label for unlabeled data by picking up the class which shows a maximum posterior probability in the prediction by itself~\cite{Lee2013a}. This method is equivalent to Entropy Regularization, which is to separate probabilistic distributions for each class by minimizing conditional entropy of the class probabilities~\cite{Grandvalet2004, Grandvalet2006}. The Entropy Regularization favors a low-density separation between classes which is a principal assumption for semi-supervised learning~\cite{Chapelle2005}. With this background, the self-training with pseudo label for the best class was applied to object detection and hyperspectral image classification~\cite{Rosenberg2005,Dopido2013}. Instead of picking up the best class, a pseudo label was estimated based on probabilistic expectation of potential labels due to the multiple target scenario for SED task~\cite{Park2021a}. Recently, Wei, et al studied a theoretical background of the empirical successes in self-training with deep learning~\cite{Wei2021}.

\subsection{Semi-supervised learning in sound event detection}
The DCASE challenge provides a framework of semi-supervised learning for sound event detection: Both the baseline implementation and the training database including unlabeled data. In the recent DCASE2020 challenge, the baseline was implemented with a Convolutional Recurrent Neural Network (CRNN) and MT model for network architecture and semi-supervised learning method respectively~\cite{Turpault2020}. The training database was composed of three subsets: a strong labeled dataset $S$, a weakly labeled dataset $W$, and an unlabeled dataset $U$. The strong label describes target sound class as well as time boundaries for each target sound interval. For the weakly labeled data, label describe a list of target sound classes without time boundaries. Unlabeled data has neither the sound class nor the time boundary. With these three types of training data, the objective function~\eqref{eq:MTloss} is modified as follows: 
\begin{equation}
\begin{aligned}
\label{eq:loss_mt}
L_{MT}=&\sum_{x \in S}BCE(f_{\theta}(x), y^s) + \sum_{x \in W}BCE(E[f_{\theta}(x)], y^w) \\
+& \delta\sum_{x \in S,W,U}MSE(f_{\theta}(x), f_{\theta'}(x')). \\
\end{aligned}
\end{equation}
where $y^{s}$ and $y^{w}$ are label for strong labeled data and weakly labeled data, respectively. $E[.]$ is expectation over the time. Note that $x'$ is generated by adding Gaussian noise to $x$ with 30 dB Signal to Noise Ratio (SNR) condition.

\begin{figure}[!t]
\centering
\includegraphics[width=0.4\textwidth]{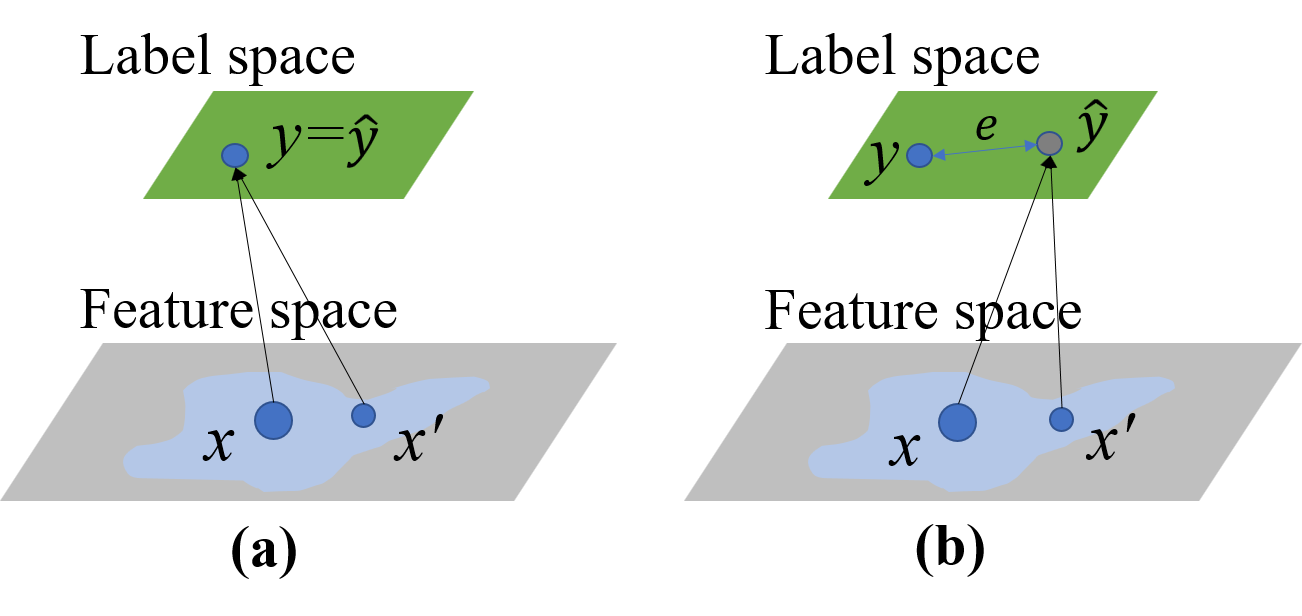}
\caption{An SED model projects inputs in feature space onto label space where $x$ and $x'$ are input features. $y$ is a true label while $\hat{y}$ is the prediction (a) Strategy of consistency-regularization method, (b) Limitation of consistency-regularization.}
\label{fig:why_pseudo}
\end{figure}

Miyazaki et al. won in the DCASE2020 challenge with an integration of strategies including a new architecture, data augmentation, classwise post-processing, and fusion~\cite{Miyazaki2020} while they employed the MT approach for semi-supervised learning. Koh, et al. suggested a Shift Consistency Training (SCT), which makes the network to produce consistent predictions for time-shifted inputs, and feature pyramid network to predict temporal label for weakly data~\cite{Koh2020}. Additionally, they explored ICT approach for SED task. With an integration system of MT, ICT, SCT, and feature pyramid, they have achieved best performance on their own. Kim, et al. proposed a modified CRNN network, which is using more filters and skip-connections with attention module, as well as modified loss function, which is a cross entropy between network prediction and pseudo label estimated by an weighted sum of predictions by the modified network and output of the challenge baseline~\cite{Kim2020, Kim2021a}. With data augmentation based on time-frequency masking and interpolation, their system outperformed the challenge baseline.

\section{Motivation of Proposed semi-supervised training method}
Fig.~\ref{fig:why_pseudo}(a) shows a concept of consistency-regularization approach. The approach allows the model producing a consistent prediction for all neighbors of one input in training set. Based on the cluster assumption, it is able to improve the performance in the classification task. In addition, it allows unlabeled data to be incorporated in supervised learning for the consistent prediction as calculated by the consistency loss. From the perspective of semi-supervised learning however, the consistency loss can be thought of as an estimation of the difference between two predictions; which in of itself could result in convergence on the wrong label as shown in Fig.~\ref{fig:why_pseudo}(b). These errors may have a great effect on the model performance since unlabeled data is generally far bigger in size than labeled data~\cite{Oliver2018}.

\begin{figure}[!t]
\centering
\includegraphics[width=0.4\textwidth]{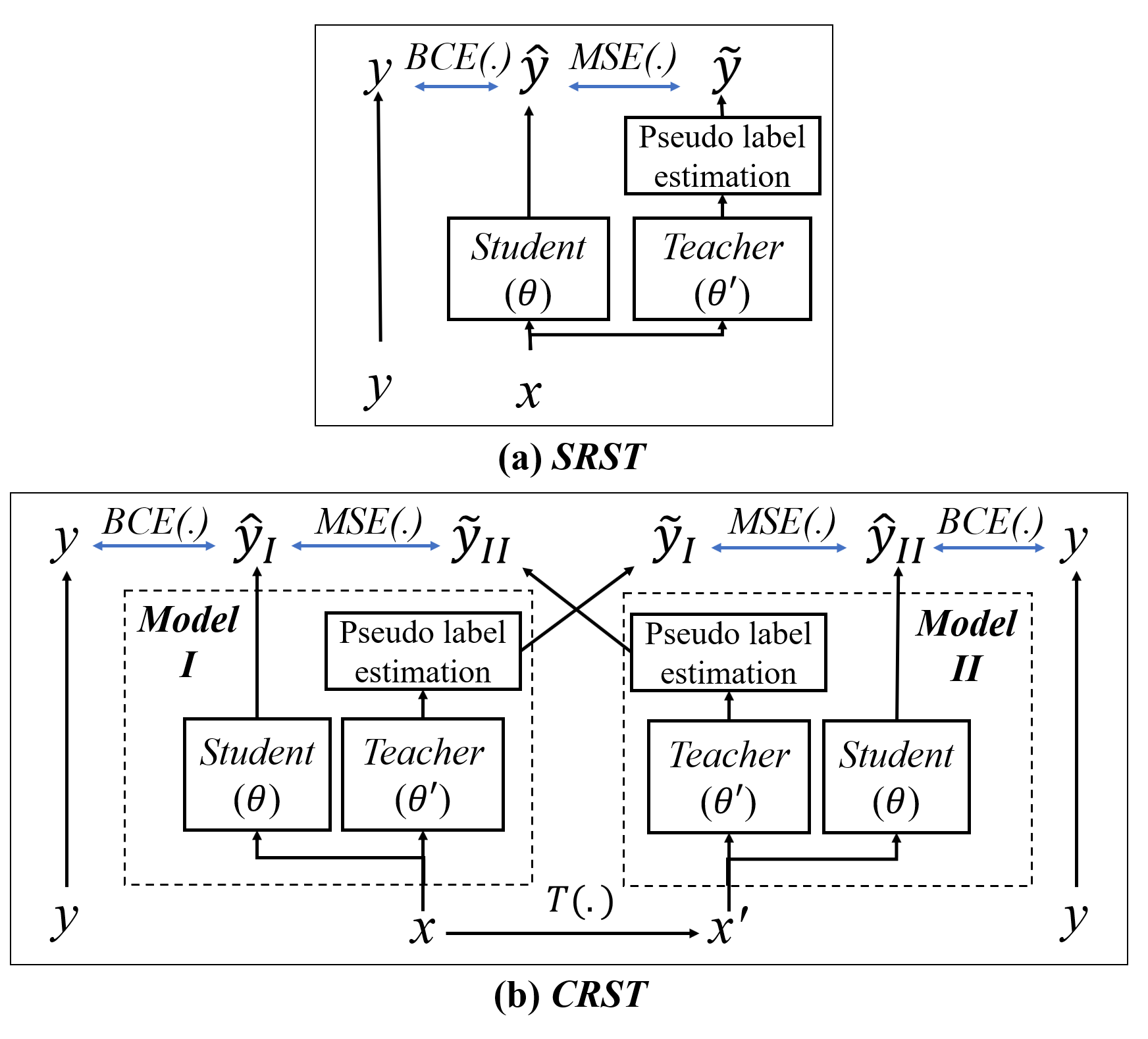}
\caption{Diagram for self-training framework where \(y\), \(\hat{y}\), and \(\tilde{y}\) is a true label, \textit{student} network's prediction, and pseudo label, respectively. \(x'\) represents manipulated data from the original \(x\) by a transformation function \(T(.)\). (a) Self-Referencing Self-Training (SRST) model, (b) Cross-Referencing Self-Training (CRST) model.}
\label{fig:diagram}
\end{figure}

In a previous work, a probabilistic expectation of potential labels was proposed as a pseudo label for unlabeled data~\cite{Park2021a}. In order to mitigate the issue of erroneous label mapping shown in Fig.~\ref{fig:why_pseudo}(b), we defined $\epsilon$ as the mean squared error between the pseudo label and network prediction, and the expectation error is then minimized. This method was shown to outperform the MT model on a SED task. However, this approach is not limitation-free, and has a risk of self-biasing because the pseudo label is estimated by itself.

As an extension of the previous work, this paper proposes a Cross-Referencing Self-Training (CRST) model to mitigate the self-biasing issue. Fig.~\ref{fig:diagram} contrasts the two approaches. Fig.~\ref{fig:diagram}a depicts the previous model, named Self-Referencing Self-Training (SRST). Fig.~\ref{fig:diagram}b depicts the proposed approach, the CRST model, which consists of two self-training models, where each model is trained on either original or manipulated data with the pseudo label estimated by the other model. The cross-referencing of a pseudo label estimated by the other model is a key point of this dual structure. It is able to avoid the self-biasing issue based on the assumption that those networks are independently trained on a different version of data. 


\section{Proposed method}
The implementation of proposed model can be found in \textit{http://github.com/JHU-LCAP/CRSTmodel}.

\subsection{Pre-processing}
An audio clip is processed to a 16kHz mono-channel audio waveform by resampling and averaging left and right channels for multi-channel audios. The audio waveform is converted to a spectrogram by performing Short Time Fourier Transform (STFT) with 2048-points frame length and 255-points hop size. Then, a log-Mel spectrogram is obtained by performing frequency integration with 128 Mel-filters spanned 0 to 8kHz frequency domain and logarithm function.
\begin{equation}
\label{eq:logmel}
\begin{aligned}
    P[n,m] &= \sum_{k}S[k,m] \times f_{mel}^{n}[k], \\
    x[n,m] &= log(max(P[n,m]^2,~\epsilon^2)),
\end{aligned}
\end{equation}
where $x$ is a log-Mel spectrogram while $S$ is a spectrogram based on STFT. $f_{mel}^{n}$ is Mel filter for the $n^{th}$ channel. $n, m$, and $k$ are indices to represent Mel filter channel, frame, and frequency bin, respectively. Note that the $\epsilon$ is set to $1.0E-5$ in order to prevent negative infinite value by the logarithm function. The different version of data, $x'$ in Fig.~\ref{fig:diagram} is generated by adding the Gaussian noise to the log-Mel spectrogram with a given Signal to Noise Ratio (SNR) condition. Note that audio length was considered up to 10 second so that zero-padding or cutting is performed for shorter or longer audios than 10 second. The SNR condition to generate manipulated data was set to 30 dB from the challenge baseline~\cite{Turpault2020}.

\subsection{Network architecture}
\label{sec:net_archi}
From the challenge baseline for SED task, CRNN is applied to both networks in \textit{Model I} and \textit{Model II}. The CRNN is composed of two stages: Convolutional Neural Network (CNN) to compress the log-Mel spectrogram into acoustic features and Bidirectional Gating Recurrent Units (BGRUs) to capture temporal relations among the compressed features by the CNN. The first stage is composed of seven convolution layers and seven averaging pooling layers. Each convolution layer uses a Gated Linear Unit (GLU) for a nonlinear activation. The GLU controls the selection of critical features in order to capture informative characteristics among the target sounds by using a self-gating function described by
\begin{equation}
\label{eq:glu}
    c' = GLU(c) = Linear(c)\times \sigma(c),
\end{equation}
where $c$ and $c'$ represents a result of convolution and GLU, respectively. $Linear(.)$ and $\sigma(.)$ is a linear transformation and a sigmoid function for the gating, respectively. In training phase, batch normalization and dropout techniques are applied to before and after performing the GLU, respectively.

The following stage consists of a Double-layered Bidirectional Gated Recurrent Unit (BGRU). The output of the CNN represents compressed features along to output channels across the time. These features are fed into the BGRU to learn the temporal characteristics of target sounds in onset and offset edges. During the training phase, dropout is applied to the end of each unit. Additionally, a linear layer with a sigmoid activation is added on the top of the BGRU to represent a presence probability of target sounds. As a result, a set of likelihood probability across the target sounds over the time is outputted by the second stage. More detailed setup for this architecture can be found in Appendix~\ref{tab:netparams} or implementation of the DCASE2020 challenge baseline for task 4~\cite{Turpault2020}.

\subsection{Objective function}
To train the \textit{student} network in each model with the three types of data, strong labeled, weakly labeled, and unlabeled, in supervised passion, the objective function consists of a classification error and an expectation error with a reliability of pseudo label. The classification error is defined by a BCE between the network prediction and strong label $y^s$ or weak label $y^w$ while a MSE between the prediction and pseudo label $\tilde{y}$ is used for the expectation error~\eqref{eq:loss}.
\begin{equation}
\begin{aligned}
\label{eq:loss}
L_{I} =& \sum_{x \in S}BCE(f_{I}(x), y^s) + \sum_{x \in W}BCE(E[f_{I}(x)], y^w) \\
+ \gamma_{II}^{s}& \sum_{x \in U}(MSE(f_{I}(x),\tilde{y}_{II})+ \gamma_{II}^{w} \sum_{x \in W}(MSE(f_{I}(x),\tilde{y}_{II}), \\
L_{II} =& \sum_{x' \in S}BCE(f_{II}(x'), y^s) + \sum_{x' \in W}BCE(E[f_{II}(x')], y^w) \\
+ \gamma_{I}^{s}& \sum_{x' \in U}(MSE(f_{II}(x'),\tilde{y}_{I})+ \gamma_{I}^{w} \sum_{x' \in W}(MSE(f_{II}(x'),\tilde{y}_{I}),
\end{aligned}
\end{equation}
where $L_{i}$ is the objective function for training \textit{Model i} whose prediction is denoted as $f_{i}(.)$. $x'$ is generated by adding Gaussian noise to original data $x$ with 30 dB Signal to Noise Ratio (SNR) condition. $E[.]$ is an averaging operator over the frames. Note that the weak label $y^w$ is a vector indicating target sound class which happened in the audio clip while the strong label $y^s$ is a matrix stacking the vectors for every frame. $\gamma_{i}^s$ and $\gamma_{i}^w$ are the reliability of pseudo label estimated in \textit{Model i} with strong labeled data and weakly labeled data, respectively. During the training phase, the parameters for both models, \textit{Model I} and \textit{Model II}, are separately optimized with each objective function~\eqref{eq:loss}.

\subsubsection{Pseudo label estimation}
Considering a scenario for multiple sound detection, a label indicating the presence of target sounds in a frame is represented to a zero vector for non-target, one-hot vector for single target, or a many-hot vector for multiple targets at least two. A pseudo label for a frame is estimated to a probabilistic expectation of those potential labels. By performing this estimation for every frame, the pseudo label of unlabeled audio clip is represented to a matrix like the strong label. Since the pseudo label has a expectation value not a binary value, the MSE criterion is more appropriate for the expectation loss in~\eqref{eq:loss} than the BCE. With this concept, the pseudo label can be estimated as
\begin{equation}
\label{eq:plabel}
    \tilde{y} = \Sigma^{K}_{k}~\Sigma^{N_{k}}_{n}~p^{k}_{n}~l^{k}_{n},
\end{equation}
where $k$ is the number of concurrent events in each frame and $n$ is an index for the case of choosing $k$-sounds of total target sounds. $l^{k}_{n}$ is a label vector expressed by a summation of delta functions like $l^{2}_{n:\{i,j\}}=\delta_{i}+\delta_{j}$ for events $i$ and $j$ ($(k=2$). $p^{k}_{n}$ is a probability of the label $l^{k}_{n}$, $K$ is maximum number of concurrent events, and $N_{k}=C!/(k! \times (C-k)!)$ is the number of potential labels under the $k$ and total number of target sound classes $C$. Note that this estimation is performed for every frame even though the frame index is omitted for brevity.

Based on the fact that the \textit{teacher} network produces more accurate predictions~\cite{Tarvainen2017}, the probability of a label vector is calculated with the \textit{teacher} prediction in each model. Based on the Bernoulli process for activation of each target sound, the probabilities $p^{k}_{n}$ are calculated depending on the number of concurrent events $k$ as
\begin{equation}
\label{eq:prob}
\begin{aligned}
    &k = 0, &p^0_{n:\{\}}&=\frac{1}{N}\Pi_{q}~(1-\hat{y}'_{q}), \\
    &k = 1, &p^1_{n:\{i\}}&=\frac{1}{N}\hat{y}'_{i}\Pi_{q \neq i}~(1-\hat{y}'_{q}), \\
    &k = 2, &p^2_{n:\{i,j\}}&=\frac{1}{N}\hat{y}'_{i}\hat{y}'_{j}\Pi_{q \neq i,j}~(1-\hat{y}'_{q}), \\
    &k = 3, &p^3_{n:\{i,j,h\}}&=\frac{1}{N}\hat{y}'_{i}\hat{y}'_{j}\hat{y}'_{h}\Pi_{q \neq i,j,h}~(1-\hat{y}'_{q}), \\
    & &...,
\end{aligned}
\end{equation}
where $N$ is a normalization factor as $N=\Sigma^{K}_{k}~\Sigma^{N_{k}}_{n}~p^{k}_{n}$. Note that the prediction by \textit{teacher} network of \textit{Model I} (\textit{Model II}) is applied to the pseudo label estimation for training \textit{Model II} (\textit{Model I}).

The estimation of a pseudo label in every frame for all unlabeled data introduces a heavy computational load in the calculation for all potential labels. To reduce this computation, the number of concurrent events $k$ is considered up to 2 based on previous work~\cite{Park2021a}. The probabilities for multi-sound labels are calculated using a dynamic programming technique~\eqref{eq:dp}.
\begin{equation}
\label{eq:dp}
\begin{aligned}
    &k = 0, &P^0&=log(p^0_{n:\{\}}),\\
    &k = 1, &P^1_{i}&=P^0 + log(\hat{y}'_{i}) - log(1-\hat{y}'_{i}), \\
    &k = 2, &P^2_{\{i,j\}} &= P^1_{i} + P^1_{j} - P^0,
\end{aligned}
\end{equation}

\subsubsection{Reliability of pseudo label}
The prediction by the \textit{teacher} network is obviously unreliable at the beginning of training. Even at later stages of training, the pseudo label is still an estimate based on the prediction. Therefore, the expectation error is weighted by the reliability of pseudo label to adjust the contribution of the error on training. In this study, the Jensen Shannon Divergence (JSD), which is bounded in [0,1], is considered to calculate the reliability $\gamma^s$ and $\gamma^w$ of the pseudo label with strong labeled and weakly labeled data, respectively~\eqref{eq:reliability}.
\begin{equation}
\label{eq:reliability}
\begin{aligned}
    \gamma^s =~& \omega \times \frac{1}{N^s}\sum_{x \in S}(1-JSD(\tilde{y}||y^s)), \\
    \gamma^w =~& \omega \times \frac{1}{N^w}\sum_{x \in W}(1-JSD(E[\tilde{y}]||y^w)), \\
    whe&re ~\omega =3.0exp(-5(1-t/T)^2), \\
    &JSD(a||b) = KLD(a||m)/2 + KLD(b||m)/2,\\
    &~~m = (a+b)/2,\\
\end{aligned}
\end{equation}
where $\omega$ is a ramp-up parameter with an index of training step $t$ and maximum number of the steps $T$, $N^s$ and $N^w$ is the number of strong labeled data and weakly labeled data, respectively. $KLD$ is a Kullback Leibler Divergence. At the beginning of training, the expectation error~\eqref{eq:loss} remains small due to the $\omega$. In later stage of training, the reliability relies on the similarity between pseudo label for labeled data and true label.

\subsection{Post-processing}
Class imbalance in training dataset is another issue often encountered in semi-supervised learning. The sparsity of the minority classes in the training set minimizes their contribution to the objective function resulting in a bias toward the majority class~\cite{Johnson2019}. Dynamic sampling~\cite{Pouyanfar2018} or data augmentation~\cite{Lee2016a} for minority class data is an effective way to resolve this issue. In the scenario of semi-supervised learning, however, it is hard to use either one because those methods need class label for all training dataset. Instead, this paper explores classwise post-processing to extract target sound intervals from class posterior probabilities over time (i.e. \textit{student} output). Once the \textit{student} model converges after training, the network's output exhibits different distribution of posteriors to each target class (Fig.~\ref{fig:post_example}). Thus, post-processing composed of thresholding and smoothing is performed with optimized classwise parameters, threshold and smoothing length.

\subsubsection{Threshold estimation} 
\begin{figure}[!t]
\centering
\includegraphics[width=0.45\textwidth]{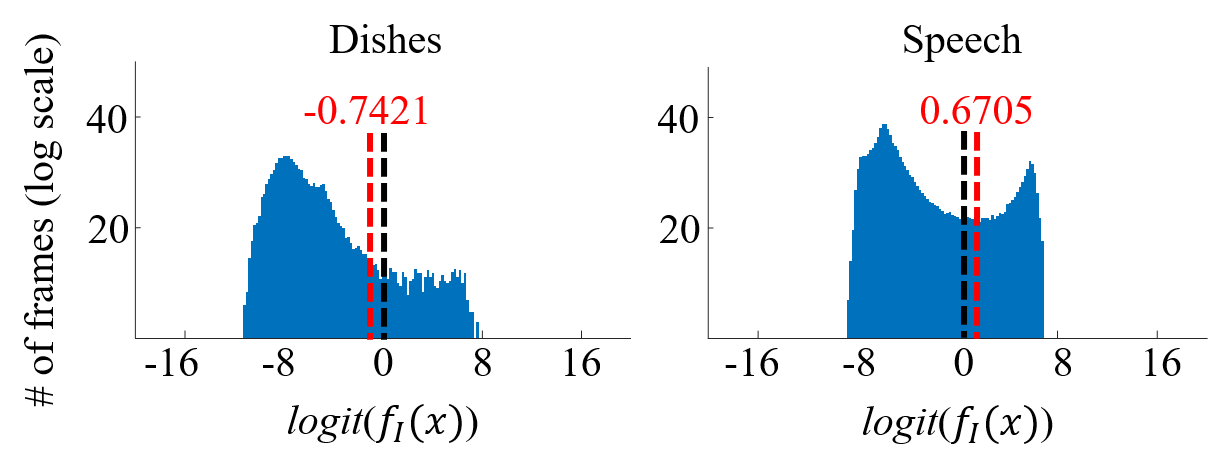}
\caption{Histograms for posterior distribution in two targets, \textit{Dishes} and \textit{Speech}. Once a model is converged in training, posteriors were calculated on weakly labeled data. Red dotted lines show optimal thresholds ($-t_{\alpha}$) for each class while black dotted line (at 0.0 in logit domain) represents a global threshold for all targets.}
\label{fig:post_example}
\end{figure}

In each target class, a threshold is estimated based on the Extreme Value Theory (EVT)~\cite{Broadwater2010}. Once a network's training converges, samples used in threshold estimation are collected by applying logit function, $logit(x)=log(\frac{x}{1-x})$, to the network prediction responding to weakly labeled data. In a threshold estimation for "\textit{Speech}" class, for example, audio clips which have the "\textit{Speech}" sounds in weakly labeled dataset are used to collect network predictions for the "\textit{Speech}" class. Since a target sound duration is typically shorter than 10 second, the samples can be categorized into two clusters for target and non-target (Fig.~\ref{fig:post_example}). These clusters are separated based on the Expectation and Maximization (EM) clustering~\cite{Yang2012a}. Threshold estimation is performed with samples belonging to the target cluster, which has a greater mean value than the other since the network has been trained. To apply the EVT to the samples, the target samples are reversed by multiplying -1. Motivated in~\cite{Gencay2001}, Cumulative Distribution Function (CDF) of the reversed samples, $F(x)$, is defined as
\begin{equation}
\label{eq:cdf}
\begin{aligned}
F(x) = (1-Pr({x \leq u}))F_{u}(x-u) + Pr({x \leq u}),\\
Pr({x \leq u})=\frac{N-n}{N},
\end{aligned}
\end{equation}
where $u$ is a predefined threshold to extract extreme samples which are greater than the predefined threshold, $Pr(x \leq u)$ is a probability of a set of samples $x$, which are less than $u$. $N$ is the total number of samples and $n$ is the number of extreme samples. In this study, $u$ was defined to the value that satisfies $Pr(x \leq u)=0.9$. Tail distribution, i.e. CDF of the extreme samples $F_{u}(x-u)$, is modelled with a Generalized Pareto Distribution (GPD)~\cite{Kang2017} as 
\begin{equation}
\label{eq:gpd}
G(z) = 1-(1+c\frac{z}{a})^{-1/c},
\end{equation}
where $G(z)$ is a CDF of the extreme samples with $z=x-u$. $a$ and $c$ are tuning parameters optimized to maximize log-likelihood for all extreme samples
\begin{equation}
\label{eq:mle}
\sum_{i=1}^{n}{log(g(z_i))} = \\ -nlog(a)-\frac{1+c}{c}\sum_{i=1}^{n}log(1+c\frac{z_i}{a}),
\end{equation}
where $g(z)$ is a probability density function of $G(z)$. Note that Nelder-Mead Simplex method~\cite{Lagarias1998} is applied to find optimal $a$ and $c$. With the parameters in \eqref{eq:cdf} and \eqref{eq:gpd}, a threshold with a given parameter $\alpha$ is estimated as 
\begin{equation}
\label{eq:threshold}
t_{\alpha}=u+\frac{a}{c}((\frac{N\alpha}{n})^{-c}-1),
\end{equation}
where $\alpha$ means a theoretical probability of false negative. Because the $t_{\alpha}$ in \eqref{eq:threshold} is derived for reversed samples in logit domain, the threshold applied to post-processing is finally obtained by $\sigma(-t_{\alpha})$.

For instance, Fig.~\ref{fig:post_example} shows histograms of posterior probabilities calculated on weakly labeled data and optimal thresholds in two classes. \textit{Dishes} sounds have an issue of imbalance between target and non-target frames because events are too short compared to whole duration. Thus, the distributions are biased towards negative values which increases likelihood to belong to a non-target frame. In this case, the threshold (marked with a black line) needs to be moved to left-side in order to enhance detection rate. On the other hand, \textit{Speech} seems to be free from this issue as shown in its distribution which is nearly zero-centered. The threshold estimation method produces a small positive shift of the global threshold, which enables to reduce false positive.

\subsubsection{Smoothing with median filter}
The smoothing is performed by applying median filter to the frames detected in the previous step. It is needed to optimize the filter length for a precise time interval since the filter length directly affects on the times at rising and falling edges. In this work, the filter length is determined by $\beta$ \% of average of sound duration. The average of sound duration is estimated with the frames resulted in thresholding with respect to weakly labeled data.

\section{Experiment}
\subsection{Database}
In order to evaluate the proposed system, the DESED database was used~\cite{Turpault2019}. This database contained 10-target sounds for SED task: \textit{Alarm\_bell\_ringing}, \textit{Blender}, \textit{Cat}, \textit{Dishes}, \textit{Dog}, \textit{Electric\_shaver\_toothbrush}, \textit{Frying}, \textit{Running\_water}, \textit{Speech}, and \textit{Vacuum\_cleaner}. The database included a training set composed of \textit{Synthetic:training}, \textit{Real:weakly labeled}, and \textit{Real:unlabeled} as shown in TABLE~\ref{tab:dataset}. To synthesize strong labeled data, background sounds were extracted from SINS~\cite{Dekkers2017}, MUSAN~\cite{Snyder2015}, or Youtube; and target sounds were obtained from freesound~\cite{Font2013}. In this training set, two points are worth highlighting: 1) Synthetic data is considered as strong labeled data instead of marking ground truth on real recordings. 2) The number of unlabeled data is much larger than the number of labeled data. For the post-processing technique proposed in this study, statistics were collected with weakly labeled data. And, two subsets, \textit{Real: validation} and \textit{Real: public evaluation} were used for performance assessment.

\begin{table}[!t]
\caption{Database}
\label{tab:dataset}
\centering
\begin{tabular}{|c|c c|}
\hline
DESED set & label type  & \# of audio clips \\
\hline
Synthetic:training & strong labeled & 2,595 \\
Real:weakly labeled & weakly labeled & 1,578 \\
Real:unlabeled & unlabeled & 14,412 \\
Real:validation & strong labeled & 1,168 \\
Real:public evaluation & strong labeled & 692 \\
\hline
\end{tabular}
\end{table}

\subsection{Evaluation criterion}
The assessment was performed with a f-score, which is a harmonic mean of precision and recall. The precision was calculated as the ratio of true-positive intervals to total detected intervals. The recall was equal to a detection rate, the ratio of true-positive intervals to target intervals. Those measures were calculated for each class. Based on the evaluation protocol for a SED task in the DCASE2020 challenge~\cite{Mesaros2016}, in this work, a detected interval was considered as true-positive if the interval satisfies three conditions: 1) onset time of the interval precedes earlier than the truth as less than a 200 ms. 2) offset time of the interval delays the truth as less than 200 ms. And 3) a sound class of the interval should be matched to the true class. Note that the 200 ms margin in both time boundaries was considered to prevent slicing the sound in the middle.

\begin{table*}[ht!]
\caption{Performance assessment with class averaging f-score}
\label{tab:overall_result}
\centering
\begin{tabular}{|c c |c c |c c|}
\hline
 & & \multicolumn{2}{|c|}{Validation} & \multicolumn{2}{|c|}{Public evaluation} \\
\hline
 & & Global & Classwise & Global & Classwise \\
\hline
Supervised learning & Strong labeled only & 17.65 $\pm$ 0.70 & & 25.15 $\pm$ 1.07 & \\
& Strong \& Weakly labeled & 28.67 $\pm$ 2.82 & & 31.32 $\pm$ 3.51 & \\
\hline
Consistency Regularization & MT (DCASE baseline) & 34.04 $\pm$ 1.47 & 36.37 $\pm$ 1.18 & 38.77 $\pm$ 1.59 & 41.09 $\pm$ 2.37 \\
& ICT (our implementation) & 35.34 $\pm$ 1.94 & 38.60 $\pm$ 1.99 & 37.19 $\pm$ 3.06 & 39.17 $\pm$ 1.90 \\
\hline
Self-training & SRST~\cite{Park2021a} & 36.36 $\pm$ 0.76 & 37.65 $\pm$ 0.91 & 37.96 $\pm$ 1.58 & 39.55 $\pm$ 1.58 \\
& SRST + aug. & 35.28 $\pm$ 1.93 & & 36.21 $\pm$ 1.08 &\\
& CRST & \textbf{37.89 $\pm$ 1.01} & \textbf{39.76 $\pm$ 1.90} & \textbf{40.19 $\pm$ 1.59} & \textbf{42.81 $\pm$ 2.03} \\
\hline
\end{tabular}
\end{table*}

\subsection{Models for SED task}
To demonstrate the effectiveness of the proposed method, CRNN network (See, Section~\ref{sec:net_archi}) was trained in different ways using a 5 times cross-validation method.

\subsubsection{Supervised learning}
This approach was considered separately for strong-labeled data only, or both strong and weakly labeled data. In the first model, only strong labeled data (synthetic data) was used for training. The loss function was defined as $L=\Sigma_{x \in S}BCE(f_{\theta}(x), y^s)$. In the second model, both strong and weakly labeled data were used for traninig. The loss function was defined as $L=\Sigma_{x \in S}BCE(f_{\theta}(x), y^s) + \Sigma_{x \in W}BCE(E[f_{\theta}(x)], y^w)$.

\subsubsection{Consistency regularization}
As a state of the art in semi-supervised learning, MT and ICT models were considered in this category. For training a MT model, the DCASE2020 challenge baseline whose objective function is denoted in~\eqref{eq:loss_mt} was used. With this implementation, ICT model was trained with modified objective function as
\begin{equation}
\begin{aligned}
\label{eq:loss_ict}
L_{ICT}= & \sum_{x \in S}BCE(f_{\theta}(x), y^s) + \sum_{x \in W}BCE(E[f_{\theta}(x)], y^w) \\
+\delta & MSE(f_{\theta}(Mix_{\lambda}(x_1, x_2)), Mix_{\lambda}(f_{\theta'}(x_1),f_{\theta'}(x_2))). \\
\end{aligned}
\end{equation}

\subsubsection{Self-referencing self-training}
The previous version of self-training, SRST model, was evaluated as well. The loss function was set to
\begin{equation}
\begin{aligned}
\label{eq:loss_st}
L_{SRST} &= \sum_{x \in S}BCE(f_{\theta}(x), y^s) + \sum_{x \in W}BCE(E[f_{\theta}(x)], y^w) \\
+& \gamma \sum_{x \in W,U}(MSE(f_{\theta}(x),\tilde{y}), \\
& where~~\gamma = min(\omega/BCE(y^s,\tilde{y}), 5.0), \\
\end{aligned}
\end{equation}
where $\tilde{y}$ is pseudo label estimated by itself and $\omega$ is defined in~\eqref{eq:reliability}. And another model (SRST+aug.) was trained with data augmentation based on adding Gaussian noise with 30 dB SNR condition. Finally, the cross-referencing self-training approach, CRST model, was evaluated.

\begin{figure*}[t!]
\centering
\includegraphics[width=1.0\textwidth]{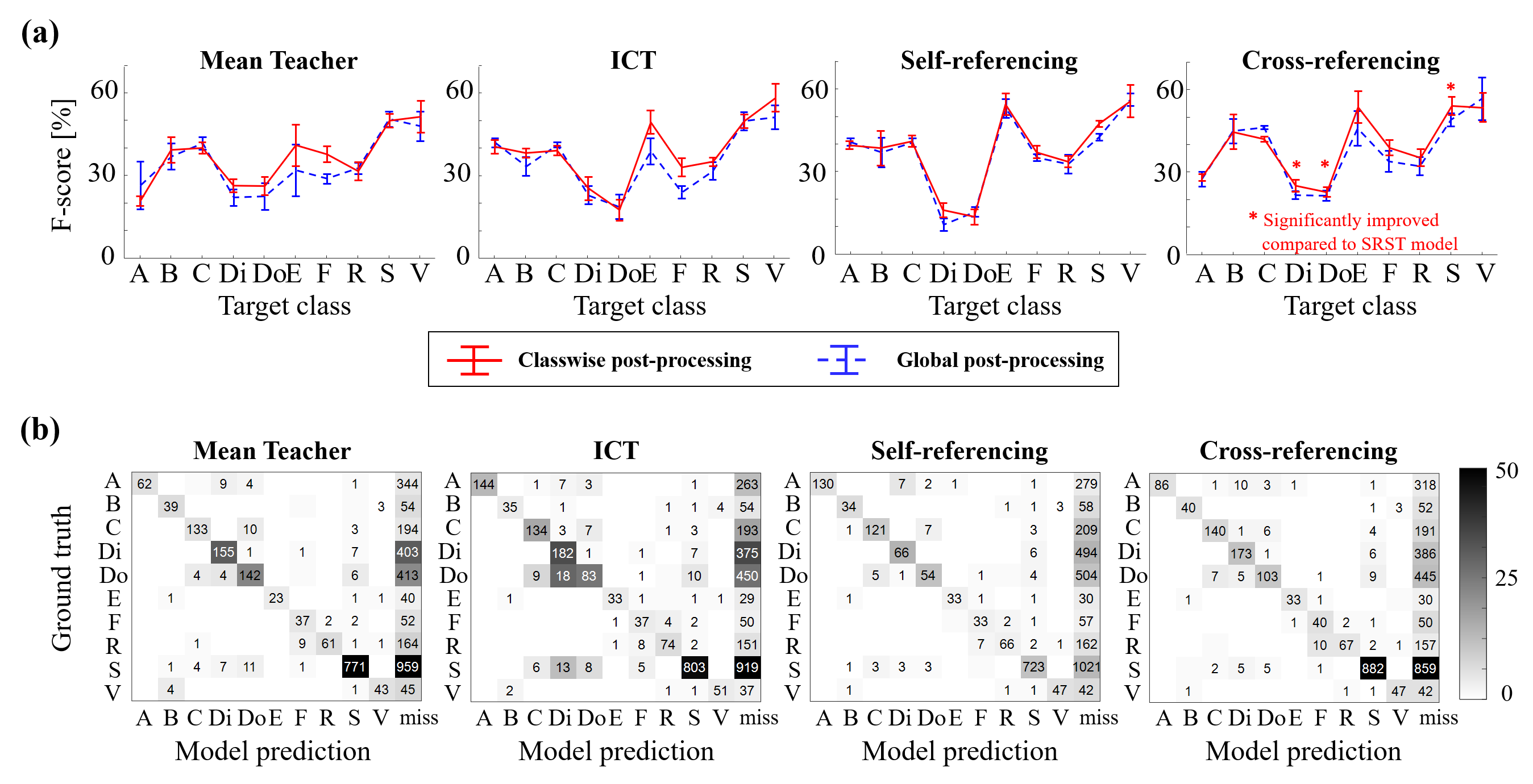}
\caption{Classwise performance on "validation" set. (a) Classwise fscores of two post-processing methods in four semi-supervised models. The results of classwise post-processing are marked as red solid line while blue dotted line is for the results of global post-processing. The error bar means the standard deviation over the 5 times repetition. (b) The most left table shows the number of sound intervals on "validation" set for each class. Then, four matrices show a confusion in classification for detected intervals which are matched to the truth in time boundaries. The numbers are the mean over the 5 times repetition and the standard deviation is represented to background light in black and white.}
\label{fig:classwise_comparison_val}
\end{figure*}

\subsection{Performance in class-averaging f-score}
TABLE~\ref{tab:overall_result} shows class averaging f-scores for each model which are the mean and standard deviation over the 5 times repetition. To investigate the effect of post-processing, f-scores obtained using the global threshold (0.5) and median filter (445ms) for all targets are summarized in the column of "Global". Note that the global parameters were determined based on the challenge baseline. Results in "Classwise" on "validation" were obtained by performing the proposed post-processing with the best parameters that were heuristically determined in searching within the intervals from 0.0002 to 0.1 with 10 steps in log-scale for $\alpha$ in~\eqref{eq:threshold} and from 5\% to 100\% with 20 steps in linear scale for $\beta$ in the estimation of filter length. Then, these best parameters were applied to the test on "public evaluation" set for each model.

Since the strong labeled dataset is composed of synthetic data produced by mixing target sounds and a background sound, it could contain artifacts such as unnatural transition in target sound boundaries and unnatural causality among target sounds. In training the synthetic data, a model could rely on these artifacts to detect target sounds. The results for two supervised models in Table~\ref{tab:overall_result} suggest that this issue could be alleviated by including real data (weakly labeled data) in training. The table also demonstrates that using unlabeled data in network training is effective to enhance the SED performance. In the evaluation on "validation" set, self-training methods except "SRST+aug." outperform the methods of consistency regularization if global post-processing is used. With 5\% significant in Welch's t-test~\cite{Derrick2017}, particularly, the CRST model improves the f-score significantly compared to MT model (\textit{p-value}=0.0013), ICT model (\textit{p-value}=0.0313), and the SRST model (\textit{p-value}=0.0265). If the classwise post-processing is performed, all of the f-scores are improved about 2.0-3.0\% compared to the results in the column of "Global". The Welch's t-test on these results confirms that the CRST model shows a significant improvement compared to MT model (\textit{p-value}=0.0096) while the CRST model averagely outperforms other two models as shown in the table (in t-test with SRST model: \textit{p-value}=0.0566, with ICT model: \textit{p-value}=0.3788).

In the second evaluation on the "public evaluation" set, the CRST model outperforms other models as well. In this evaluation, the MT model shows the second best in the average of f-scores over the repetition. In T-test with the results of global post-processing, the p-values are 0.1931, 0.0869, and 0.0560 in comparison to the MT model, the ICT model, and the SRST model, respectively. If the classwise post-processing is performed, the CRST model shows a significant improvement compared to the ICT model (\textit{p-value}=0.0192) and the SRST model (\textit{p-value}=0.0221). On the other hand, the CRST model averagely outperforms than the MT model (\textit{p-value}=0.2536).

From these evaluations, the CRST model results in more accurate detection of sound intervals with stable performance on both datasets. Additionally, the proposed post-processing enables to improve the performance about 2.0-3.0\% in class-averaging f-score.

\subsection{Investigation of classwise performance}
To explore f-scores in individual class, classwise f-scores for semi-supervised models are represented in Fig.~\ref{fig:classwise_comparison_val}. In the classwise comparison on "validation" set (Fig.~\ref{fig:classwise_comparison_val}(a), results of classwise post-processing), the CRST model has reached the best performance for five classes and the second best for two classes in the mean of 5 times repetition. With 5\% significant in the T-test, particularly, the model shows a significant improvement in \textit{Electric\_shaver\_toothbrush} and \textit{Speech} compared to the MT model. Compared to the ICT model, significant improvement could be found in \textit{Dog} and \textit{Speech}. And \textit{Dishes}, \textit{Dog}, and \textit{Speech} are significantly improved compared to the SRST model. In a classwise performance, the SRST model shows the biggest variation across the target classes among the four models. The f-scores of the SRST model are averagely better than other models in \textit{Alarm\_bell\_ringing} and \textit{Electric\_shaver\_toothbrush}. Among the four models, the worst f-score for \textit{Dishes} and \textit{Dog} could be found in the SRST model as well. It is difficult to investigate what happen exactly in these classes during the training due to the lack of labels in the training data. One of potential reasons is that the SRST model was biased by a pseudo label that was estimated incorrectly due to some reasons such as a noise or an overlapping effect. Once the model miss \textit{Dishes} sounds, the model never detect the \textit{Dishes} sounds because the pseudo label is unable to give any information for the \textit{Dishes}. Therefore, the \textit{Dishes} class would be getting worse and worse, on the other hand, the \textit{Electric\_shaver\_toothbrush} class would be getting better and better because it is able to reflect more samples for the \textit{Electric\_shaver\_toothbrush} class in training. This issue has been resolved with the cross-referencing framework as shown in the result of the CRST model. To investigate the detection results of each model, each confusion matrix for the results based on classwise post-processing is represented on Fig.~\ref{fig:classwise_comparison_val}(b). With detected intervals that have matched to the truth in time boundaries, the confusion matrix was built by counting the number of the intervals in classification. To give the information about missing intervals in time boundaries, the most right column in each confusion matrix shows the number of the missing intervals by the model. Note that the numbers on each confusion matrix are the mean value over the repetition and the standard deviation is represented to background light. In both the ICT and SRST model, the f-score in \textit{Dog} is the worst among the targets. In case of the ICT model, the worst result is owing to a confusion among the classes. On the other hand, inaccurate detection in time is the reason of the SRST model as in that only 65 intervals were detected with correct time boundaries.

\begin{figure*}[t!]
\centering
\includegraphics[width=1.0\textwidth]{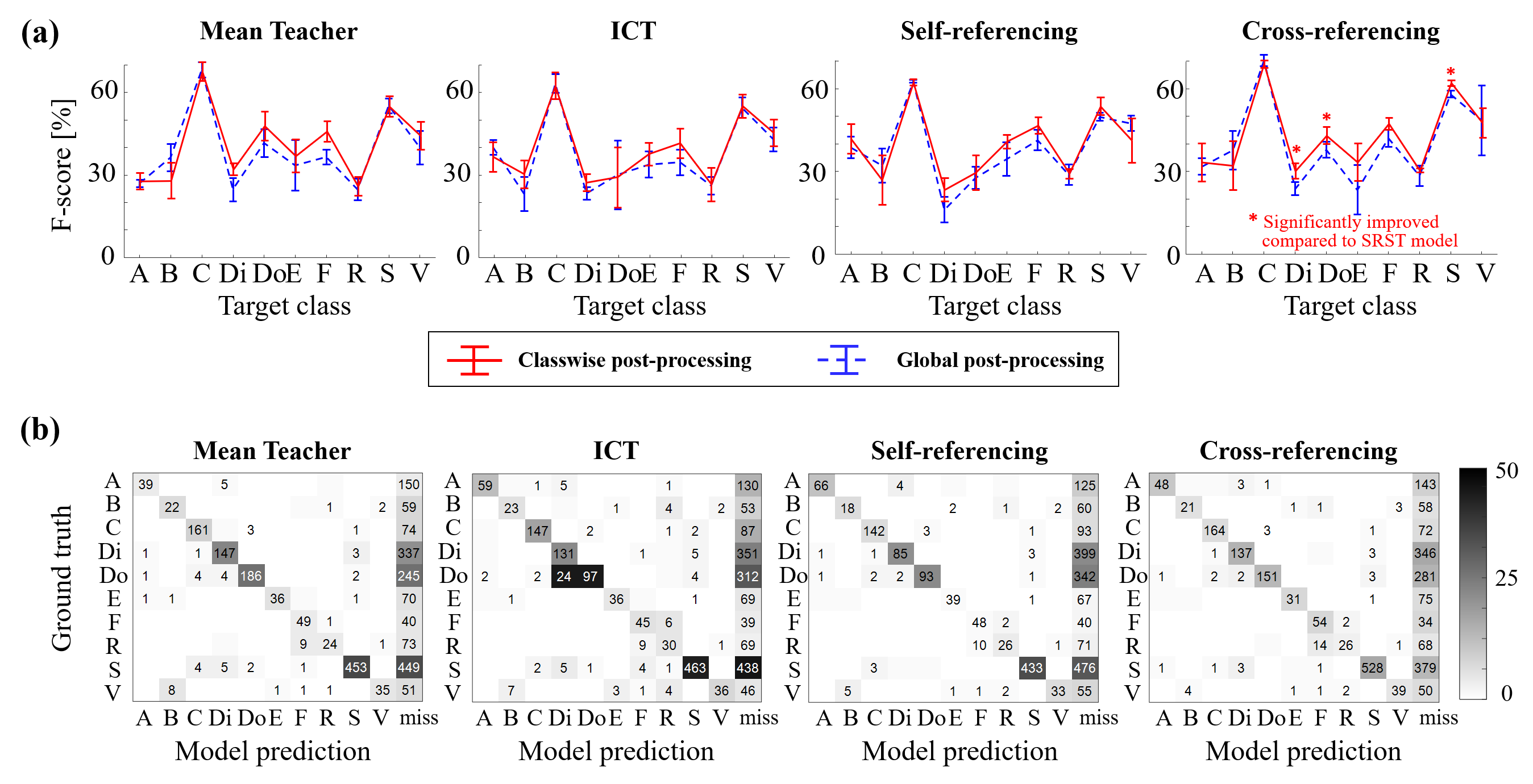}
\caption{Classwise performance on "public evaluation" set. (a) Classwise fscores of two post-processing methods in four semi-supervised models. The results of classwise post-processing are marked as red solid line while blue dotted line is for the results of global post-processing. The error bar means the standard deviation over the 5 times repetition. (b) The most left table shows the number of sound intervals on "validation" set for each class. Then, four matrices show a confusion in classification for detected intervals which are matched to the truth in time boundaries. The numbers are the mean over the 5 times repetition and the standard deviation is represented to background light in black and white.}
\label{fig:classwise_comparison_eval}
\end{figure*}

The results performed on "public evaluation" set are summarized in Fig.~\ref{fig:classwise_comparison_eval}. As shown in Fig.~\ref{fig:classwise_comparison_eval}(a), the CRST model reached the best performance in six classes and the second best in two classes. In this assessment, the MT model outperforms the ICT and SRST model as shown in Table~\ref{tab:overall_result}. In a classwise comparison, the ICT and SRST model outperforms the MT and CRST models in \textit{Alarm\_bell\_ringing} while these models still have in trouble to detect \textit{Dishes} and \textit{Dog} compared to other two models. And the CRST model shows a significant improvement in \textit{Speech} compared to all other models. Similarly, the confusion matrices in this assessment are represented in Fig.~\ref{fig:classwise_comparison_eval}(b). The trend is mostly consistent with the results on "validation" set except in \textit{Cat}, \textit{Dog} and \textit{Electric\_shaver\_toothbrush}. The sounds of "Cat" and "Dog" have a big variation depending on species, size, or age, which could explain the different performance between the "validation" and "public evaluation" sets. It can also be noted that the number of \textit{Cat} and \textit{Dog} intervals is less than the number of the sounds in "validation set". On the other hand, more sounds for \textit{Electric\_shaver\_toothbrush} are included in the "public evaluation" set, and the performance in this category shows worse than the f-score in "public evaluation" set. Note that the total number of target intervals in this evaluation can be found by a summation of each row of the confusion matrix.

\begin{table}[ht!]
\caption{Percentage of the number of concurrent sound \\ intervals in both test datasets}
\label{tab:concurrent}
\centering
\begin{tabular}{|c|c c c|c c c|}
\hline
\multicolumn{1}{|c|}{} & \multicolumn{3}{c|}{Validation} & \multicolumn{3}{c|}{Public evaluation} \\
\hline
 & $k=1$ & $k=2$ & $k>2$ & $k=1$ & $k=2$ & $k>2$ \\
\hline
A & 82.14  & 17.38 & 0.48 & 67.35 & 30.10 & 2.55 \\
B & 61.46 & 38.54 & 0 & 72.62 & 26.19 & 1.19 \\
C & 89.44 & 10.56 & 0 & 87.08 & 11.67 & 1.25 \\
Di & 41.09 & 48.15 & 10.76 & 37.91 & 56.56 & 5.53 \\
Do & 81.23 & 18.6 & 0.17 & 77.33 & 21.54 & 1.13 \\
E & 46.15 & 53.85 & 0 & 44.44 & 54.63 & 0.93 \\
F & 11.7 & 61.7 & 26.6 & 24.44 & 58.89 & 16.67 \\
R & 65.4 & 31.65 & 2.95 & 67.89 & 30.28 & 1.83 \\
S & 55.25 & 42.13 & 2.62 & 41.62 & 54.66 & 3.72 \\
V & 69.57 & 30.43 & 0 & 82.29 & 17.71 & 0 \\
\hline
Total & 62.18 & 34.47 & 3.35 & 55.37 & 41.27 & 3.36 \\
\hline
\end{tabular}
\end{table}

\subsection{Exploration of maximum number of concurrent events}
In order to reduce a computational load in pseudo label estimation, in this study, the number of concurrent events, $k$, was considered up to 2 ($K=2$) so that total 56 potential labels ($=1+10+45$ for $k=0$, $k=1$, and $k=2$, respectively) were used to estimate pseudo label. According to the previous study in SRST model~\cite{Park2021a}, the class averaging f-score was saturated at $K=2$ because the case, which three or more target sounds happen at a time, is unusual in practical environments. With both test datasets, sound intervals that are overlaid with other target sound were counted and the results are summarized in TABLE~\ref{tab:concurrent}. In total, the ratio of cases of three or more concurrent sounds, is less than 5 \% in both datasets. Thus, the preset parameter for the maximum number of concurrent sounds is acceptable assumption in the pseudo label estimation.

\subsection{Effect of preset parameters in post-processing}
\begin{figure}[!t]
\centering
\includegraphics[width=0.45\textwidth]{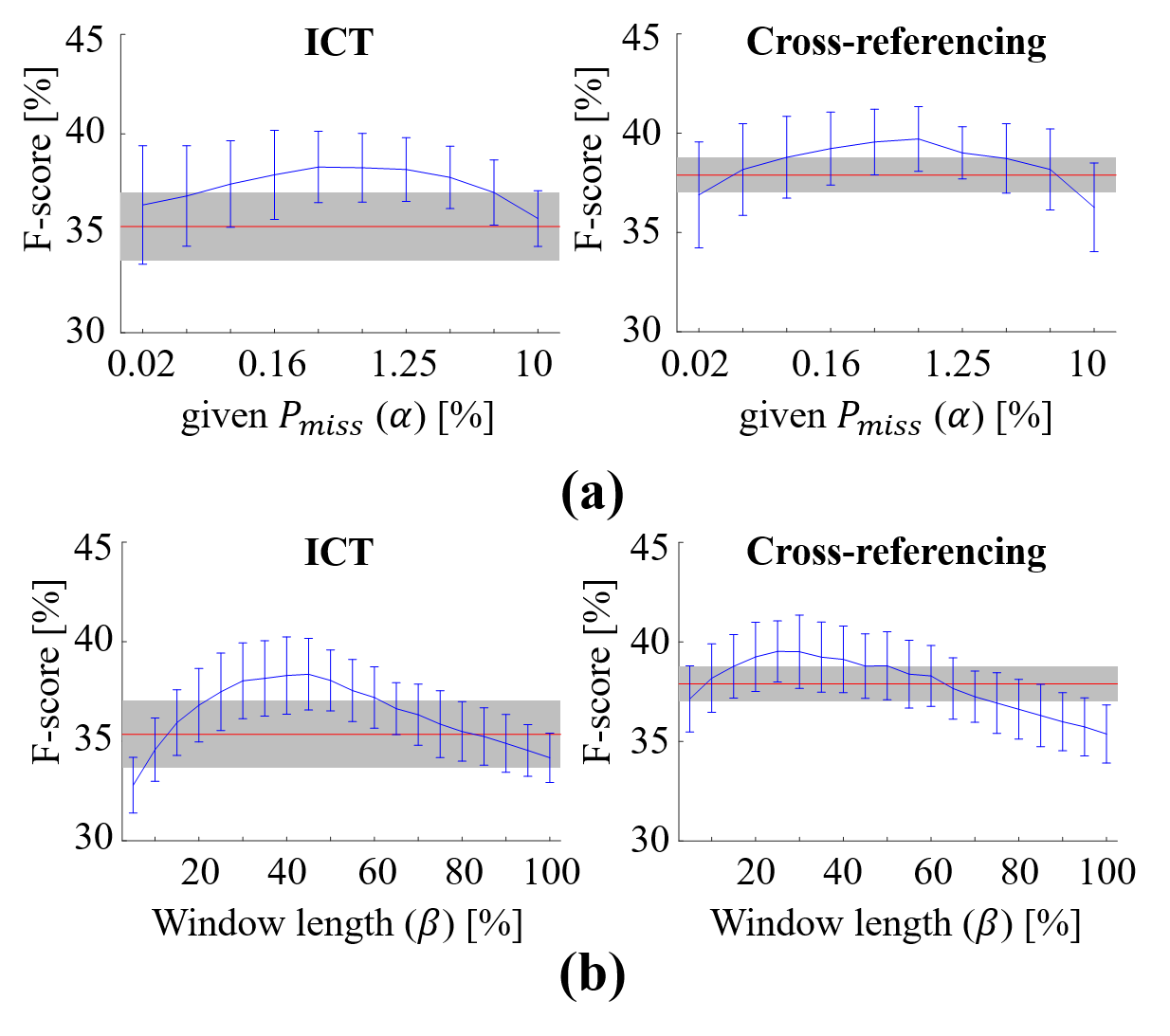}
\caption{Class averaging f-score based on classwise post-processing with different preset parameters. The red line on each panel represents the fscores based on global post-processing and the gray region and error bar represent a 95 \% confidence intervals for the mean value. (a) depending on parameter $\alpha$ in ICT and CRST model with $\beta=45\%$ and $\beta=25\%$. (b) depending on parameter $\beta$ in ICT and CRST model with $\alpha=0.32\%$ and $\alpha=0.64\%$.}
\label{fig:effect_postprocessing}
\end{figure}

As shown in the results (Fig.~\ref{fig:classwise_comparison_val}(a) and \ref{fig:classwise_comparison_eval}(a)), the performance is mostly improved by applying the classwise post-processing compared to the results of the global post-processing. For performing the classwise post-processing, preset parameters, a theoretical false negative rate $\alpha$ and a ratio to the average of sound duration $\beta$, are heuristically decided in searching within the intervals from 0.02\% to 10\% with 10 steps in log-scale for $\alpha$ in~\eqref{eq:threshold} and from 5\% to 100\% with 20 steps in linear scale for $\beta$ in the estimation of filter length. For the CRST and ICT (the second best model on "validation" set as in TABLE~\ref{tab:overall_result}), class averaging f-scores depending on the parameters are represented in Fig.~\ref{fig:effect_postprocessing}. Fig.~\ref{fig:effect_postprocessing}(a) shows class averaging f-score depending on theoretical false negative rate \(\alpha\) in \eqref{eq:threshold}. In both panels, the red line represents the result based on the global post-processing. The gray region and the error-bar represent a 95 \% confidence interval for the mean. Note that the $\beta$ was set to 45\% and 25\% for ICT and CRST model, respectively. According to the results, the f-score has been improved significantly on 0.32 and 0.64 \% for both models. If the $\alpha$ is too big or small, it is unable to improve the performance due to the false results such as FPs or FNs. The f-scores depending on $\beta$ are represented in Fig.~\ref{fig:effect_postprocessing}(b). Note that the $\alpha$ was set to 0.32\% and 0.64\% for ICT and CRST model. As shown in results, the length of smoothing filter is a critical parameter effecting on the performance since the smoothing makes an early or a delay on time boundaries of detected intervals. If a short length filter is used in the smoothing, it is unable to remove very short intervals resulted in the thresholding. Thus, the smoothing remains the FPs, short-time intervals due to a noise, and makes small precision in the evaluation. On the other hand, it could merge two intervals which are close to each other in time when a long length filter is used. In this case, the smoothing could reduce TPs and make FPs more because the merging interval would be decided to FPs due to the mismatching in time boundaries.

\begin{table}[!t]
\caption{Class averaging f-score in different perturbations}
\label{tab:perturb}
\centering
\begin{tabular}{|c|c c|}
\hline
\multicolumn{1}{|c|}{perturbation} & \multicolumn{2}{c|}{Validation} \\
& Global & Classwise \\
\hline
Adding noise & 37.89 $\pm$ 1.01 & 39.76 $\pm$ 1.90 \\
Mixup & 34.35 $\pm$ 1.40 & 37.21 $\pm$ 0.79 \\
Shifting & 39.26 $\pm$ 1.77 & 41.61 $\pm$ 0.79 \\
\hline
\hline
\multicolumn{1}{|c|}{perturbation} & \multicolumn{2}{c|}{Public evaluation} \\
& Global & Classwise \\
\hline
Adding noise & 40.19 $\pm$ 1.59 & 42.81 $\pm$ 2.03 \\
Mixup & 36.15 $\pm$ 1.10 & 40.35 $\pm$ 0.90 \\
Shifting & 40.48 $\pm$ 1.53 & 43.26 $\pm$ 1.79 \\
\hline
\end{tabular}
\end{table}

\subsection{Effect of data perturbation}
For evaluation of the proposed method, perturbation was produced by adding Gaussian noise to original data. To explore an effect of perturbation, additional evaluations were performed with other perturbation methods such as mixup~\cite{Zhang2018} and frame-shift. For data mixup, $Mix_{\lambda}(a,b)$ defined in~\eqref{eq:ICTloss} was used to generate manipulated data for two original data $a$ and $b$. A delay factor (i.e. the number of frame) for frame shifting was generated by Gaussian random process with zero-mean and 40-standard deviation. Then, a new data point was generated by truncating 10 second from the delay factor and padding with the rest of original data. The results are summarized in TABLE~\ref{tab:perturb} with the mean and standard deviation in 5 cross-validation repetitions. 

As noted in the results, frame-shifting is the most effective technique tested. Adding Gaussian noise is also comparable to the frame-shifting (\textit{p-value}=0.0799 and \textit{p-value}=0.7215 in validation and public evaluation for classwise post-processing, respectively). On the other hand, the mixup method appears to yield worst results. One explanation is that this technique generates more overlapping events which may bias \textit{Model II} to learn mapping of target sound features from these overlapping sounds. Naturally, another limitation is that the pseudo label estimation assumes a cap of 2 on the number of concurrent events, informed by the statistics of real test audios as shown in TABLE~\ref{tab:concurrent}. 

\section{Discussion}
Current state-of-the-art systems for sound event detection have been leveraging a combination of approach such as data augmentation, network architecture, semi-supervised learning, post-processing, and fusion in order to yield the best performance on the DCAS challenge. One of the best performing systems proposed by Miyazaki et al. explores all these aspects~\cite{Miyazaki2020}. A self-attention model based on either transformer~\cite{Kong2020} or conformer~\cite{Gulati2020} was used instead of the CRNN. And the self-attention model was trained based on MT approach. Although the system finally reached to 50.6\% for class averaging f-score, it seems that the new model is comparable to the challenge baseline. When the self-attention model is trained without data augmentation and tested without both post-processing and fusion, the performance was reported to 28.6 \% and 34.4 \% for transformer and conformer, respectively. Although the f-scores were respectively improved to 41.0\% and 41.7\% by performing their classwise post-processing, it has an issue for a generalization of the parameters such threshold and filter length because both parameters were manually optimized on the "validation" set.

Koh, et al. proposed a SCT loss and a deeper network based on modified CRNN, where a pooling is only performed on frequency domain for a high time-resolution, and feature-pyramid (FP) that leverages to discard unreliable predictions is incorporated with the CRNN~\cite{Koh2020}. In post-processing, the target sounds were grouped into two groups, background sounds and impulsive sounds, depending on the sound duration, then two different filters, whose length was heuristically decided, were applied for the sounds. According to the technical report, their implementation for ICT shows better performance than the ICT result in TABLE~\ref{tab:overall_result}. A possible reason is the use of ICT loss in combination with the MT loss. Also interpolation and time-frequency shifting were considered as a part of data augmentation from the loss function for SCT. The data augmentation in this manner seems be effective in improving the performance as shown in other submissions~\cite{Miyazaki2020, Yao2020, Liu2020, Kim2020, Huang2020a}. Since this study focuses on a method for semi-supervised learning, a simple way like adding noise is only considered to make manipulated data samples $x'$ in Fig.~\ref{fig:diagram}.

Kim, et al. proposed another modified CRNN which is using more channels and skip-connected convolution layers with attention modules~\cite{Kim2020}. Instead of MT approach, the network was trained with a cross-entropy between model prediction and pseudo label to resolve the issue of MT approach in Fig.~\ref{fig:why_pseudo}(b). The pseudo label was estimated on an weighted sum of model prediction and truth label with a preset weight. According to the technical report, it seems that this approach outperformed the results in TABLE~\ref{tab:overall_result}. However, it is unfair to directly compare to the numbers in the TABLE~\ref{tab:overall_result} because Kim's method used a different architecture in the number of channels and skip connection as well as data augmentation based on interpolation and time-frequency masking.

In this study, all other configurations such as perturbation (i.e. data augmentation) and network architecture were fixed because this study tackles to develop an advanced method for semi-supervised learning. The effect of other configuration on the SED performance will be investigated as a future work.

\section{Conclusion}
Sound event detection aims to identify sounds of interest as sound category and time boundaries for each sound interval. Deep neural networks are powerful models for the sound event detection; however, they are faced with the challenge of data acquisition and curation in order to provide accurate estimate of events that can guide supervised training of these networks. The current paper explores self-training using cross-referencing as well as classwise post-processing in order to leverage the unlabeled training data. The cross-referencing self-training is composed of two separate models. In training, one model estimates a pseudo label for unlabeled data with the prediction by itself, then the other model refers the pseudo label to train itself vice versa. In this way, these two models are separately trained so that a self-biasing risk in self-referencing model could be resolved. Additionally, a post-processing composed of thresholding and smoothing is explored to find sound intervals from the model prediction. This paper introduces a threshold estimation based on Extreme Value Theory and filter length estimation with the statistics of model prediction. These proposed approaches are tested on sound event detection task described by the recent DCASE challenge and shown to result in improved performance compared to the state-of-the art in semi-supervised learning. 


%

\appendices
\section{Network architecture of CRNN and Training parameters}
\label{tab:netparams}
The basic network, Convolutional Recurrent Neural Network (CRNN), consists of 7-convolutional blocks for CNN and 2-Bidirectional Gate Recurrent Unit (BGRU) for RNN. Each convolutional block was composed of 2D-convolution layer with [$3 \times 3$] kernel and [$1 \times 1$] stride, 2D batch normalization layer, activation layer, dropout layer, and average pooling layer. The number of kernels was set to [16, 32, 64, 128, 128, 128 ,128] per block. The 2D batch normalization was performed along to the output channels. Gated Linear Unit (GLU) in~\eqref{eq:glu} was applied to the activation layer. The dropout parameter was set to 0.5. In the average pooing layer, the stride size was defined as [[2, 2], [2, 2], [1, 2], [1, 2], [1, 2], [1, 2], [1, 2]] per block. Note that the pooling was performed twice along to time axis while the pooling was performed in 7-times along to frequency axis. The number of nodes in BGRU was set to 128 as the number of kernels at the top of the previous CNN.

Each model on this structure was trained for 300 epochs with Adam optimizer. And the best model on a cross-validation set with global post-processing is kept. The remaining parameters defined for training are following: 24 batch size (6 strong labeled data, 12 unlabeled data, and 6 weakly labeled data), adaptive learning rate limited to 0.001 for maximum. 

\section*{Acknowledgment}
This work was supported by NIH U01AG058532, ONR N00014-17-1-2736, N00014-19-1-2689 and NSF 1734744.

\ifCLASSOPTIONcaptionsoff
  \newpage
\fi

\begin{IEEEbiography}[{\includegraphics[width=1in,height=1.25in,clip,keepaspectratio]{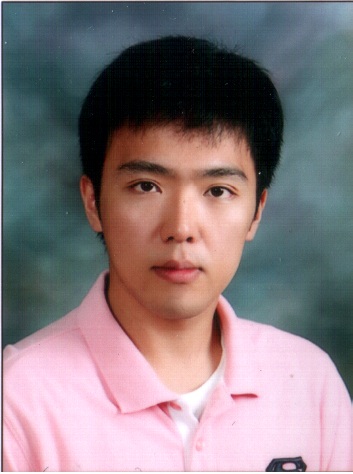}}]{Sangwook Park} received the B.S. degree in Electrical and Electronics Engineering from Chung-Ang University, Seoul, South Korea in 2012, and Ph.D. degree in electrical and computer engineering from Korea University, Seoul, South Korea in Aug. 2017. After a year of serving as Research Professor for audio signal analysis at Korea University, he has been PostDoc fellow of Electrical and Computer Engineering in Johns Hopkins University since Sept. 2018. His current works involve acoustic signal analysis for detection and classification and building a computational model emulating mammalian auditory system.
\end{IEEEbiography}

\begin{IEEEbiography}[{\includegraphics[width=1in,height=1.25in,clip,keepaspectratio]{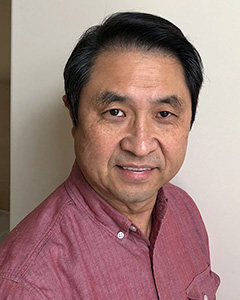}}]{David K. Han} (Senior Member, IEEE) is the inaugural holder of the Bruce Eisenstein Endowed Chair and Professor of Electrical and Computer Engineering at Drexel University. He is an ASME Fellow and an IEEE senior member. He has previously held positions as research and faculty member at the Johns Hopkins University Applied Physics Laboratory, Army Research Laboratory, and the University of Maryland at College Park, and has additionally served as Distinguished IWS Chair Professor at the US Naval Academy. Han has authored or coauthored over 100 peer-reviewed papers, including four book chapters. His current research interests include computer vision, speech recognition,machine learning, and robotics.
\end{IEEEbiography}


\begin{IEEEbiography}[{\includegraphics[width=1in,height=1.25in,clip,keepaspectratio]{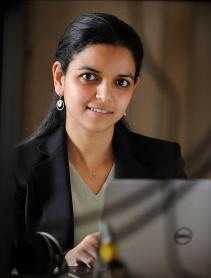}}]{Mounya Elhilali}
(Senior Member, IEEE) received her Ph.D. degree in electrical and computer engi- neering from the University of Maryland, College Park, MD, USA. She is a professor of Electrical and Computer Engineering with a secondary appointment at the department of Psychology and Brain Science at Johns Hopkins University. She directs the Laboratory for Computational Audio Perception, which examines human and machine hearing, with a focus on robust representation of sensory information in noisy soundscapes, problems of auditory scene analysis and cognitive control of auditory perception. She was named the Charles Renn faculty scholar and is the recipient of the National Science Foundation CAREER award and the Office of Naval Research Young Investigator award.
\end{IEEEbiography}




\end{document}